\documentclass[aps,pra,groupedaddress,twocolumn,showpacs]{revtex4-1}
\usepackage[latin1]{inputenc}
\usepackage{amssymb}
\usepackage{graphicx}
\usepackage{pstricks}
\usepackage{amsmath}

\begin{document}
\title{Negative index of refraction in a four-level system with magnetoelectric cross coupling and local field corrections}

\author{F. Bello${}$}

\affiliation{${}$Departament de F$\acute{\imath}$sica, Universitat Aut$\grave{o}$noma de Barcelona, E-08193 Bellaterra, Spain}

\date{\today{}}

\begin{abstract}
This research focuses on a coherently driven four-level atomic medium with the aim of inducing a negative index of refraction while taking into consideration local field corrections as well as magnetoelectric cross coupling, i.\noindent e.\noindent\;chirality, within the material's response functions. Two control fields are used to render the medium transparent for a probe field which simultaenously couples to an electric and a magnetic dipole transition, thus allowing one to test the permittivity and permeability of the material at the same time. Numerical simulations show that a negative index of refraction with low absorption can be obtained for a range of probe detunings while depending on number density and the ratio between the intensities of the control fields.
\end{abstract}
\maketitle

\section{Introduction}\label{sec1}

The search for a negative refractive index within the visible range of frequencies has attracted a considerable amount of interest due in part to the advances made in the production of optical metamaterials (see for instance \cite{shalaev07} and references therein), where the optical properties of the medium are determined mainly by its structure rather than by its intrinsic properties. Materials with a negative index of refraction, also known as left-handed materials, have a negative phase velocity with power flow in the opposing direction, and have been theoretically proven to exhibit phenomena such as subwavelength resolution, a reverse Doppler effect, and Cherenkov radiation to name a few \cite{veselago68,pendry00}. Although experiments have verified the existence of negative index materials and a wide range of applications \cite{shelby01,smith00,cai09}, losses remain a major limitation in the optical regime \cite{xiao09,Wuestner10,xiao10}. Alternatively, it was shown recently that negative refractive index can be electromagnetically induced in vapors with low absorption rates \cite{oktel04,thommen06} thus eliminating the need of manufacturing the medium. In addition, recent publications \cite{kastel07c,kastel07,orth08,kastel09,zhang08,li09,yavuz10} have pointed out that if one considers magnetoelectric cross coupling and/or local field corrections, a negative index of refraction can be achieved without requiring a negative permeability, which so far has been one of the largest impediments to overcome in the field.  

Using density matrix theory in order to treat the interaction of light with matter semiclassically, this paper studies a dense medium of 4-level atoms interacting with two intense pump fields which serve to modify the response of a weak probe field. The probe field is coupled simultaneously to both an electric and a magnetic dipole transition. Included are local field corrections (LFCs) and chirality, which is explicitly shown to help produce a negative refractive index as well as tune the system between an amplifying (parametric) and absorbing medium. 

In section \ref{sec2} the theory needed to calculate the index of refraction in a four-level system is introduced which includes local field effects and magnetoelectric cross coupling. The numerical results for a dense atomic gas of neon are presented in section \ref{sec3} showing the possibility of obtaining a negative refractive index with low absorption. The paper concludes in section \ref{conclusions}.

\section{Theoretical Modeling}\label{sec2}

Fig.~1 shows the 4-level system under consideration that was originally introduced by Thommen \textit{et al.} \cite{thommen06}.  Although here, the magnetic component of the probe field ($\vec{B}$) is explicitly included which couples to transition $|1\rangle \leftrightarrow |2\rangle$ with Rabi frequency $\Omega_{B}=\vec{m}_{12}\cdot\vec{B}/\hbar$ where $\vec{m}_{12}$ is the corresponding magnetic dipole moment. The electric component of the probe field ($\vec{E}$) couples to transition $|3 \rangle \leftrightarrow |4\rangle$ with Rabi frequency $\Omega_{E}=\vec{d}_{34}\cdot\vec{E}/\hbar$, where $\vec{d}_{34}$ is the electric dipole moment. Both transitions are nearly degenerate, allowing the probe field with frequency $\omega_{12}^L(\equiv \omega_{34}^L)$ to simultaneously test both the permeability and permittivity of the system. Additionally, two control lasers with frequencies $\omega_{13}^L$ and $\omega_{24}^L$ induce the magnetoelectric cross coupling that must be taken into consideration when solving for the material's response functions. The two control beams couple to transitions $|1\rangle\leftrightarrow|3\rangle$ and $|2\rangle\leftrightarrow|4\rangle$ with Rabi frequencies $\Omega_{13}$ and $\Omega_{24}$, respectively. The system's detunings are defined as $\delta_{ij}=\omega_{ij}^L-\left|(\epsilon_{i}-\epsilon_{j})/\hbar\right|$ for which $i,j=1,2,3,4$ and $i\neq j$. $\epsilon_{i}$ denotes the energy of the corresponding atomic level. 

In the rotating wave approximation, the Hamiltonian describing the 4-level system is written as,
\begin{eqnarray}\label{Eq:22}
H &=& \sum\limits_{i=1}^{4}\epsilon_{i}\sigma_{ii} + \frac{1}{2}\{-\Omega_{B}e^{i\omega_{12}^Lt}\sigma_{12}- \hbar \Omega_{13}e^{i\omega_{13}^Lt}\sigma_{13}\nonumber \\&-& \hbar \Omega_{24}e^{i\omega_{24}^Lt}\sigma_{24}- \Omega_{E}e^{i\omega_{12}^Lt}\sigma_{34}+h.c.\},
\end{eqnarray}
where $\sigma_{ij}=|i\rangle \langle j|$ ($i,j\in 1\rightarrow 4$) are level shift operators. Given the closed loop scheme presented, at least one of the Rabi frequencies must be allowed to be complex \cite{morigi02}.

\begin{figure}
	\centering
	\includegraphics{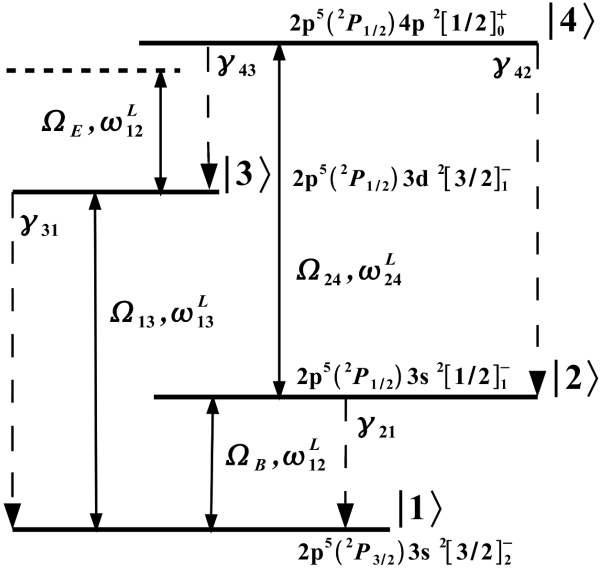}
	\caption{4-level system under consideration. A probe field with frequency $\omega_{12}^L$ couples the electric ($|3\rangle\leftrightarrow|4\rangle$) and magnetic ($|1\rangle\leftrightarrow|2\rangle$) transitions with Rabi frequencies $\Omega_{E}$ and $\Omega_{B}$, respectively. Two intense fields with Rabi frequencies $\Omega_{13}$ and $\Omega_{24}$ couple the optical transitions ($|1\rangle\leftrightarrow|3\rangle$) and ($|2\rangle\leftrightarrow|4\rangle$) with the former being a two photon transition due to parity \cite{thommen06}.  The corresponding wavelengths for Neon are $\lambda_{12}=5.4\,\mu\mbox{m}$, $\lambda_{13}=704\,\mbox{nm}$, and $\lambda_{24}=352\,\mbox{nm}$ with a relatively small energy difference of $0.16\,$MHz between the two transitions coupled to the probe field. The spontaneous decay rates are denoted by $\gamma_{21}$, $\gamma_{31}$, $\gamma_{42}$ and $\gamma_{43}$ while the energy levels are given in jL-coupling notation.}
\end{figure}

Assuming a dense atomic medium, local field corrections (LFCs) are included for which  mean-field theory is utilized \cite{friedberg89,jyotsna95,manassah00,dowling93}. The macroscopic electric fields in the Hamiltonian are therefore replaced by the microscopic, local fields according to the Lorentz-Lorenz relation given by,
\begin{equation}\label{Eq:5}
\vec{E}^{\mbox{\tiny{LFC}}}=\vec{E}+\frac{4\pi}{3}\vec{P},
\end{equation}
with $\vec{P}=\eta\langle \vec{d} \rangle $, where $\eta$ is the density of atoms and $\langle \vec{d} \rangle$ is the expectation value of the electric dipole moment under consideration. Similarly, in order to consider LFCs on the magnetic field one can write an analogous equation to that above:
\begin{equation}\label{Eq:8}
\vec{H}^{\mbox{\tiny{LFC}}}=\vec{H}+\frac{4\pi}{3}\vec{M},
\end{equation}
with $\vec{M}=\eta\langle \vec{m} \rangle $, and $\langle\vec{m}\rangle$ yielding the expectation value of the magnetic moment. The external magnetic field is taken to be $\vec{H}=\vec{B}=\pm i\vec{E}$ with the minus or plus sign referring to $\sigma^{+}$ or $\sigma^{-}$ circularly polarized light, respectively.

A general equation for the polarization or magnetization of a medium should include cross-terms in order to account for the possibility of chirality, i.e., part of the polarization (magnetization) is induced by the magnetic (electric) field. In order to calculate the system's permeability and permittivity in terms of the electric, $\alpha_{\mbox{\tiny{E}}}$, and magnetic $\alpha_{\mbox{\tiny{B}}}$ polarizabilities, the theory previously introduced by K$\ddot{\mbox{a}}$stel \textit{et al}. \cite{kastel09} is followed with the equations for the polarization and magnetization given by,
\begin{eqnarray}\label{Eq:10}
\vec{P} &=& \eta\alpha_{\mbox{\tiny{E}}}\vec{E}^{\mbox{\tiny{LFC}}} + \eta\alpha_{\mbox{\tiny{EH}}}\vec{H}^{\mbox{\tiny{LFC}}},\\  
\vec{M} &=& \eta\alpha_{\mbox{\tiny{HE}}}\vec{E}^{\mbox{\tiny{LFC}}} + \eta\alpha_{\mbox{\tiny{H}}}\vec{H}^{\mbox{\tiny{LFC}}},
\end{eqnarray}
where $\alpha_{\mbox{\tiny{EH}}}$ and $\alpha_{\mbox{\tiny{HE}}}$ are defined as the cross-coupling polarizabilities due to chirality. 

The permittivity ($\varepsilon$) and permeability ($\mu$) are defined in relation to the susceptibilities such that $\varepsilon=1+4\pi\chi_{e}$ and $\mu=1+4\pi\chi_{m}$ with $\chi_{e}(\chi_{m})$ being the electric (magnetic) susceptibility of the system.  Assuming the material's overall response to be isotropic, one then makes use of the general definitions for polarization and magnetization to calculate the susceptibilities and eventually the index of refraction. Hence, within isotropic chiral media we have the definitions,
\begin{eqnarray}\label{Eq:17}
\vec{P} &=& \chi_{e}\vec{E}+\frac{\xi_{\mbox{\tiny{EH}}}}{4\pi}\vec{H},\\
\vec{M} &=& \frac{\xi_{\mbox{\tiny{HE}}}}{4\pi}\vec{E}+\chi_{m}\vec{H},
\end{eqnarray}
with $\xi_{\mbox{\tiny{EH}}}=4\pi\frac{\eta}{\kappa}\alpha_{\mbox{\tiny{EH}}}$ and $\xi_{\mbox{\tiny{HE}}}=4\pi\frac{\eta}{\kappa}\alpha_{\mbox{\tiny{HE}}}$ \cite{kastel09}. Using Eqs.~(4) and (2) (Eqs.~(5) and (3)), one is able to obtain the permittivity and permeability in terms of polarizabilities, yielding
\begin{equation}\label{Eq:24.1}
\varepsilon=1+4\pi\frac{\eta}{\kappa}\times\{\alpha_{\mbox{\tiny{E}}}+\frac{4\pi}{3}\eta(\alpha_{\mbox{\tiny{EH}}}\alpha_{\mbox{\tiny{HE}}}-\alpha_{\mbox{\tiny{H}}}\alpha_{\mbox{\tiny{E}}})\},
\end{equation}
\begin{equation}\label{Eq:25}
\mu=1+4\pi\frac{\eta}{\kappa}\times\{\alpha_{\mbox{\tiny{H}}}+\frac{4\pi}{3}\eta(\alpha_{\mbox{\tiny{EH}}}\alpha_{\mbox{\tiny{HE}}}-\alpha_{\mbox{\tiny{H}}}\alpha_{\mbox{\tiny{E}}})\}
\end{equation}
with 
\begin{eqnarray}\label{Eq:26}
\kappa=1&-&\frac{4\pi}{3}\eta\alpha_{\mbox{\tiny{E}}}-\frac{4\pi}{3}\eta\alpha_{\mbox{\tiny{H}}}\\\nonumber &-&(\frac{4\pi}{3})^{2}\eta^{2}\times(\alpha_{\mbox{\tiny{EH}}}\alpha_{\mbox{\tiny{HE}}}-\alpha_{\mbox{\tiny{H}}}\alpha_{\mbox{\tiny{E}}}).
\end{eqnarray}
Without the effects of chirality, the last term in parenthesis of (\ref{Eq:24.1}), (\ref{Eq:25}), and (\ref{Eq:26}) would be left out. The first two terms in (\ref{Eq:26}) that include only $\alpha_{\mbox{\tiny{H}}}$ $(\alpha_{\mbox{\tiny{E}}})$ are due to LFCs, and if chirality is not included then only the  $\alpha_{\mbox{\tiny{H}}}$ $(\alpha_{\mbox{\tiny{E}}})$ term would be added to $\mu (\varepsilon)$, however with chirality both terms appear in the denominator of the permittivity and permeability.  The index of refraction, which is now dependent on the cross-coupling polarizabilities, is derived from the Helmholtz equation and given by \cite{pendry04,kastel09}:
\begin{equation}\label{Eq:26.5}
n^{\pm}=\sqrt{\varepsilon\mu-\frac{(\xi_{EH}+\xi_{HE})^{2}}{4}}\pm\frac{i}{2}(\xi_{HE}-\xi_{EH}),
\end{equation}
\centerline{}
where $n^{\pm}$ refers to $\sigma^{\pm}$ polarization.  
Note that a negative index of refraction does not necessarily require that both the electric and magnetic responses of the material be negative at the same time.  This lessens the demand to induce a strong magnetic response which as previously mentioned has been one of the biggest obstacles to overcome \cite{kastel07,orth08,kastel09}.  

\begin{figure}
	\centering
	\includegraphics{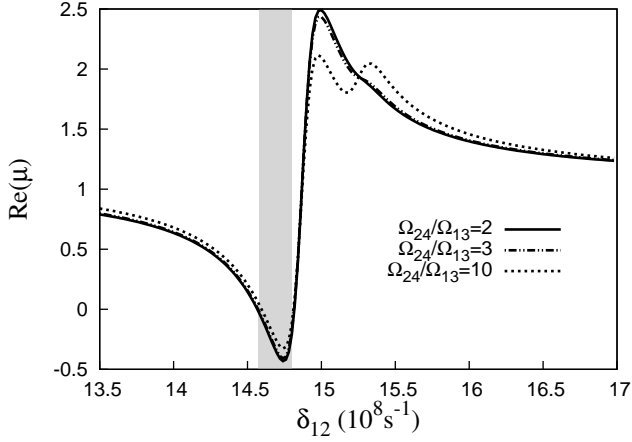}
  \caption{Shown is the real part of the permeability as a function of probe field detuning for various coupling field ratios $\Omega_{24}/\Omega_{13}=2$ (solid line), $\Omega_{24}/\Omega_{13}=3$ (dashed-dotted line), and $\Omega_{24}/\Omega_{13}=10$ (dotted line). The shaded area indicates the interval over which the real part of the permeability takes negative values. (See text for additional parameters.)}
\end{figure}

To characterize the medium's response, the ratio between the real and imaginary parts of the index of refraction ($n=n^{'}+in^{''}$) is used as a comparison between the phase velocity and the rate of absorption. Ideally, one aims for a negative real part of the index of refraction along with a low imaginary part representing low absorption. The figure of merit is defined as \cite{kastel09,orth08}:
\begin{equation}\label{Eq:25.1}
\mbox{FoM}=\left|\frac{n^{'}}{n^{''}}\right|.
\end{equation} 

\section{Numerical results}\label{sec3}

In order to solve for the dynamics of the system, the Liouville equation is numerically simulated, $\dot{\rho}= -\frac{i}{\hbar}[H,\rho]+ \Gamma \rho$, where $\Gamma \rho $ accounts any decay associated with spontaneous emission. For the four-level system under investigation the full density matrix solutions are given by,

\begin{figure}[t]
	\centering
	\includegraphics{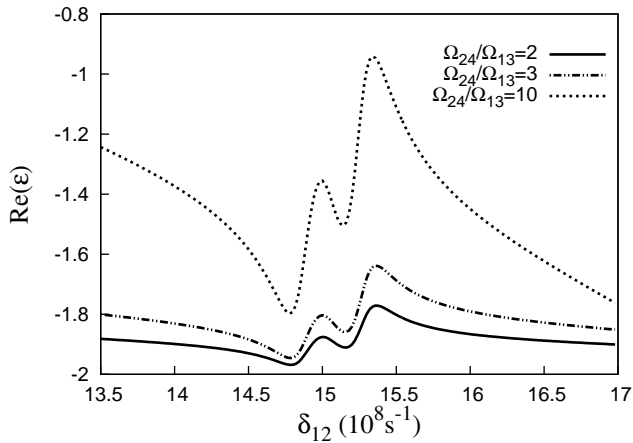}     
	\caption{The real part of the permittivity is plotted as a function of the probe detuning, $\delta_{12}$, for various ratios between coupling fields; $\Omega_{24}/\Omega_{13}=2$ (solid line), $\Omega_{24}/\Omega_{13}=3$ (dashed-dotted line), and $\Omega_{24}/\Omega_{13}=10$ (dotted line).}
\end{figure}

\begin{widetext}
\begin{eqnarray}\label{Eq:30}
\dot{\rho}_{11}&=&-2{\rm Im}\left\{(\Omega_B+L_{12}\rho_{12})\rho_{21}+(\Omega_{13}+L_{13}\rho_{13})\rho_{31}\right\}+\gamma_{21}\rho_{22}+\gamma_{31}\rho_{33} \\
\dot{\rho}_{22}&=&-2{\rm Im}\left\{(\Omega_B^*+L_{12}^*\rho_{21})\rho_{12}+(\Omega_{24}+L_{24}\rho_{24})\rho_{42}\right\}-\gamma_{21}\rho_{22}+\gamma_{42}\rho_{44} \\
\dot{\rho}_{33}&=&-2{\rm Im}\left\{(\Omega_{13}^*+L_{13}^*\rho_{31})\rho_{13}+(\Omega_{E}+L_{34}\rho_{34})\rho_{43}\right\}-\gamma_{31}\rho_{33}+\gamma_{43}\rho_{44} \\
\dot{\rho}_{12}&=&i\delta_{12}\rho_{12}+i(\Omega_B+L_{12}\rho_{12})(\rho_{22}-\rho_{11})+i(\Omega_{13}+L_{13}\rho_{13})\rho_{32}-i(\Omega_{24}^*+L_{24}^*\rho_{42})\rho_{14}-\Gamma_{12}\rho_{12}\\
\dot{\rho}_{13}&=&i\delta_{13}\rho_{13}+i(\Omega_{13}+L_{13}\rho_{13})(\rho_{33}-\rho_{11})+i(\Omega_{B}+L_{12}\rho_{12})\rho_{23}-i(\Omega_{E}^*+L_{34}^*\rho_{43})\rho_{14}-\Gamma_{13}\rho_{13}\\
\dot{\rho}_{14}&=&i\delta_{14}\rho_{14}+i(\Omega_{B}+L_{12}\rho_{12})\rho_{24}+i(\Omega_{13}+L_{13}\rho_{13})\rho_{34}-i(\Omega_{24}+L_{24}\rho_{24})\rho_{12}-i(\Omega_{E}+L_{34}\rho_{34})\rho_{13}-\Gamma_{14}\rho_{14}\,\,\\
\dot{\rho}_{23}&=&i\delta_{23}\rho_{23}+i(\Omega_{B}^*+L_{12}^*\rho_{21})\rho_{13}-i(\Omega_{13}+L_{13}\rho_{13})\rho_{21}+i(\Omega_{24}+L_{24}\rho_{24})\rho_{43}-i(\Omega_{34}^*+L_{34}^*\rho_{43})\rho_{24}-\Gamma_{23}\rho_{23}\,\,\\
\dot{\rho}_{24}&=&i\delta_{24}\rho_{24}+i(\Omega_{24}+L_{24}\rho_{24})(\rho_{44}-\rho_{22})+i(\Omega_{B}^*+L_{12}^*\rho_{21})\rho_{14}-i(\Omega_{E}+L_{34}\rho_{34})\rho_{23}-\Gamma_{24}\rho_{24}\\
\dot{\rho}_{34}&=&i\delta_{34}\rho_{34}+i(\Omega_{E}+L_{34}\rho_{34})(\rho_{44}-\rho_{33})+i(\Omega_{13}^*+L_{13}^*\rho_{31})\rho_{14}-i(\Omega_{24}+L_{24}\rho_{24})\rho_{32}-\Gamma_{34}\rho_{34}
\end{eqnarray}
\end{widetext}
where $L_{12}=\left|\vec{m}_{12}\right|^2 4\pi \eta/(3\hbar)$ and $L_{ij}=|\vec{d}_{ij}|^2 4\pi \eta/(3\hbar)$ with $ij=\{13,24,34\}$.  These terms take into account the local field corrections which are applied to all of the fields within the Hamiltonian according to the definitions given in Eqs.~(\ref{Eq:5}) and (\ref{Eq:8}). The dipole moments are taken to be real, hence $\vec{d}_{ij}=\vec{d}_{ji}$ and $\vec{m}_{12}=\vec{m}_{21}$. The decay of the off-diagonal elements of the density matrix, $\Gamma_{ij}$, are given in the radiative limit. In addition, we consider the Rabi frequencies of the coupling beams to be real while the Rabi frequencies of the probe field are taken to be complex with $\Omega_{B}=\Omega^0_{B}e^{i\phi}$ and $\Omega_{E}=\Omega^0_{E}e^{i\phi}$.  It has previously been shown that by adjusting the phase angle $\phi$, one can minimize the imaginary part of the index of refraction, hence maximizing absorption or amplification of the probe field \cite{zhang08,li09}.  For our simulations, the special case of $\phi=\pi$ is used. Also note that as seen in Eqs.~(16)-(21), the LFCs yield the following effective detuning $\delta^{\mbox{\tiny{LFC}}}_{ij}=\delta_{ij}+L_{ij}(\rho_{jj}-\rho_{ii})$, known as the Lorentz shift. 

By numerically integrating the density matrix equations, we are able calculate the material response for a wide range of parameters. The coherences between the states coupled to the probe field yield the polarization, $\vec{P}=\eta\rho_{43}\vec{d}_{34}$, and magnetization, $\vec{M}=\eta\rho_{21}\vec{m}_{12}$. 
Following Ref.~\cite{kastel09}, in order to calculate the polarizabilities from Eqs.~(4) and (5) we expand the coherences of the probe field transitions as power series in $\vec{E}^{\mbox{\tiny{LFC}}}$ and $\vec{H}^{\mbox{\tiny{LFC}}}$. From the polarizabilities the permittivity and permeability are then obtained from Eqs.~(8) and (9) and the index of refraction from Eq.~(11).

\begin{figure}
              \begin{center}
\begin{tabular}{lc}
               \;\; (a)  \\
                \centerline{\includegraphics{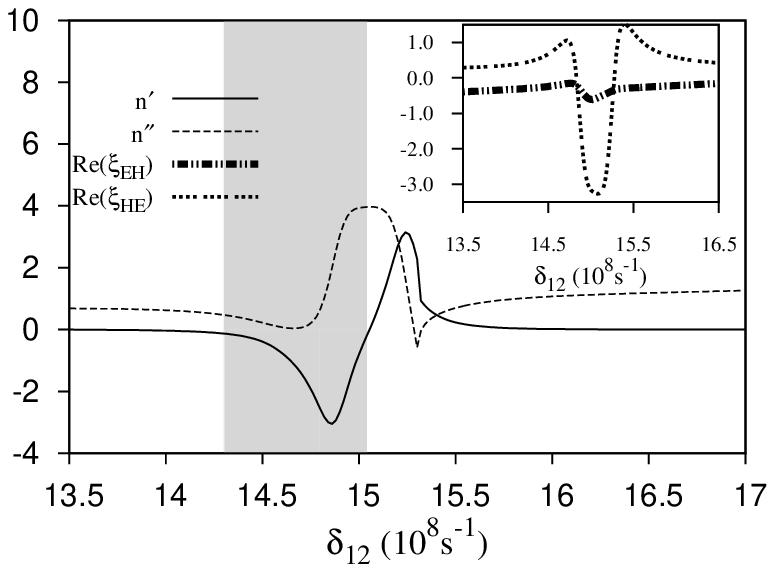}} \\
               \;\; (b)  \\
                \centerline{\includegraphics{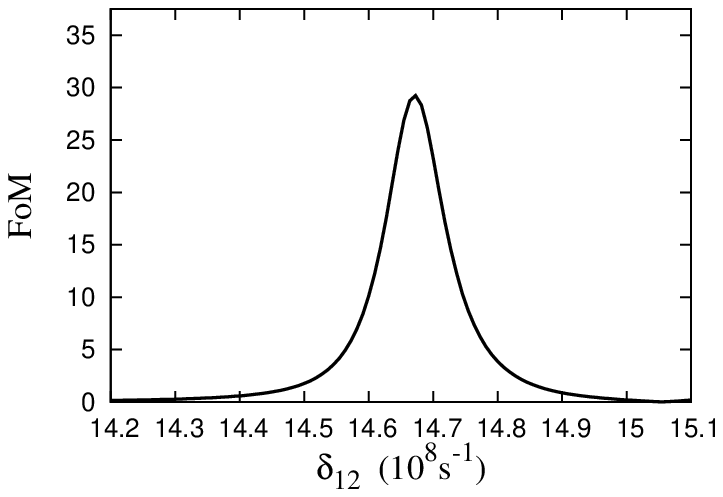}} \\
\end{tabular}
             \caption{Shown are (a) the real (solid line) and imaginary (dashed line) parts of the index of refraction, and (b) the values for the figure of merit as a function of probe detuning. A negative index of refraction is shown to roughly occur for a detuning between 14.3$\times10^{8}\,$s$^{-1}$ and 15$\times10^{8}\,$s$^{-1}$ (shaded area) with a FoM$\approx$30. Results presented are for a coupling ratio of $\Omega_{24}/\Omega_{13}=10$. The inset in (a) shows the real part of the chiral terms which contribute to the imaginary part of the index of refraction.}
\end{center}
\end{figure}

Figs.~2 and ~3 display the real part of the permeability and permitivity, respectively, for different ratios of $\Omega_{24}/\Omega_{13}$ with $\delta_{13}=-1.5\times 10^6$s$^{-1}$, $\eta=1.0\times10^{17}\,$at/cm$^{3}$,$\gamma_{31}=\gamma_{43}=\gamma_{42}=10^7\,$s$^{-1}$, and $\gamma_{21}=10^7/\alpha^{2}\,$s$^{-1}$ where $\alpha=1/137$ is the fine structure constant. The largest Rabi frequency $\Omega_{24}= 15\times10^8\,$s$^{-1}$ is kept constant throughout the simulations corresponding to a laser intensity on the order of 10$^{2}$ W/cm$^{2}$. For the case of Neon gas, this is well within the reported breakdown intensity of approximately 10$^{14}$ W/cm$^{2}$ for a density on the order of 10$^{21}$ atoms/cm${^{2}}$ if one is using a pulse of 7ps\cite{panarella76}. The strong effects of chirality begin to be seen as the ratio between coupling fields is increased and we notice the appearance of multiple peaks in the permeability and permitivity. This is due to the dependence of the cross-polarizabilities not only on the real parts but also the imaginary parts of the coherences of the probe field transitions. At very large detunings, the material's response functions (not shown on Figs.~2 and ~3) converge towards the expected values for a field travelling through free space.

Fig.~4 shows the calculated results for the index of refraction which becomes negative for an interval of $\sim 50\times 10^6\,$s$^{-1}$ (shaded area) over the detuning of the probe field with a figure of merit of roughly $30$ corresponding to a coupling ratio of $\Omega_{24}/\Omega_{13}=10$ (see Fig.~4(b)). Note that this range of probe field detuning exhibits a negative index of refraction larger than the range in which the permitivity (shaded area in Fig.~2) is negative. The real part of the two chiral terms $\xi_{EH}$ and $\xi_{HE}$ (inset) plays a crucial role in decreasing absorption as the difference between the two directly affects the imaginary part of the index of refraction (see Eq.~(\ref{Eq:26.5})). The two peaks corresponding to the Re($\xi_{HE}$) help produce the two troughs in the curve representing $n^{''}$ and yield the highest values for the figure of merit, thus providing the points for which we find the minimum amount of absorption. By adjusting the ratio between coupling fields $\Omega_{24}$ and $\Omega_{13}$, we have ourselves a useful tool with which we can further increase the influence of the second term outside the square root of Eq.~(11) in order to manipulate the index of refraction.

\begin{figure}[t]
              \begin{center}
\begin{tabular}{lc}
               \;\; (a)  \\
                \centerline{\includegraphics{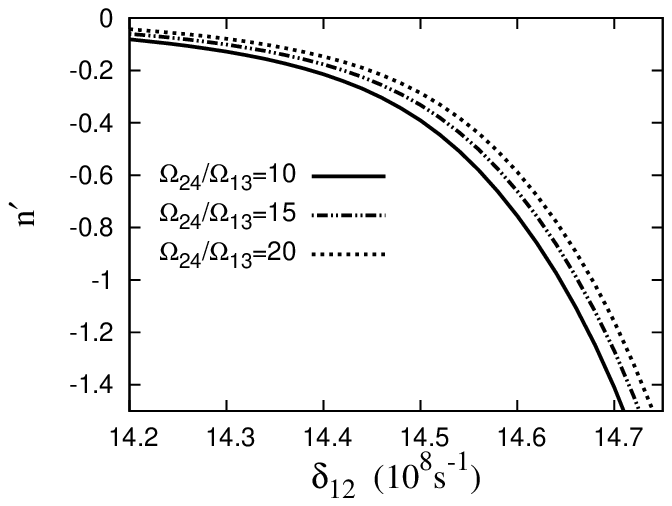}} \\
               \;\; (b)  \\
                \centerline{\includegraphics{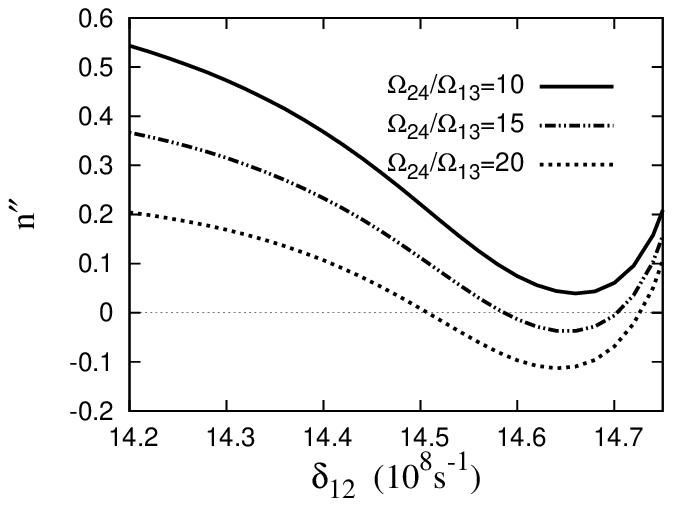}} \\
\end{tabular}
             \caption{Shown is (a) the real and (b) the imaginary part of the index of refraction as a function of the probe detuning, $\delta_{12}$, for coupling field ratios of $\Omega_{24}/\Omega_{13}=10$ (solid line), $\Omega_{24}/\Omega_{13}=15$ (dashed-dotted line), and $\Omega_{24}/\Omega_{13}=20$ (dotted line).}
\end{center}
\end{figure}

\begin{figure}[t]
              \begin{center}
\begin{tabular}{lc}
               \;\; (a)  \\
                \centerline{\includegraphics{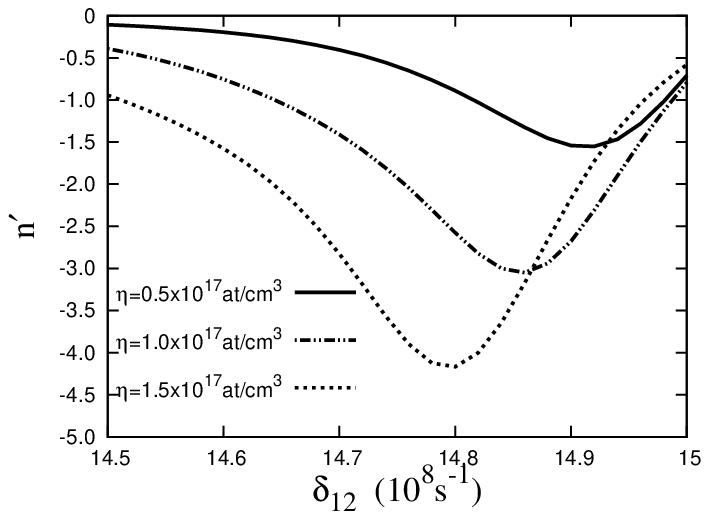}} \\
               \;\; (b)  \\
                \centerline{\includegraphics{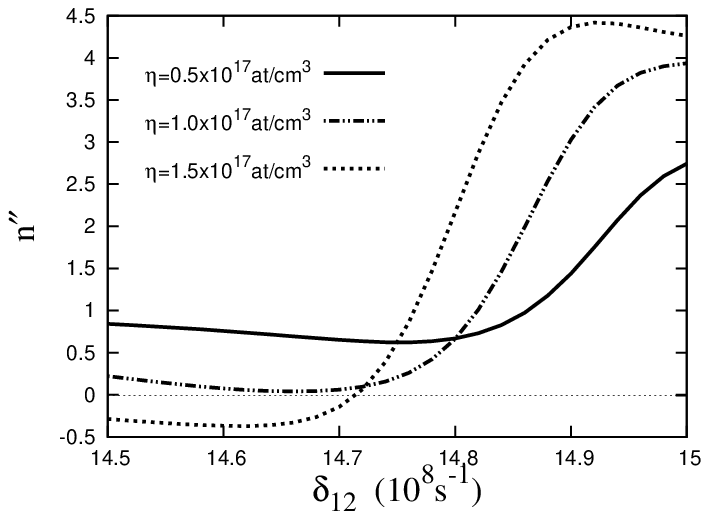}} \\
\end{tabular}
             \caption{Shown is (a) the real and (b) the imaginary part of the index of refraction as a function of the probe detuning, $\delta_{12}$, for different values of the atomic density ($\eta$). All other parameters are identical to the case for neon gas presented in Fig.~4.}
\end{center}
\end{figure}

More recently, research has focused on producing a negative refractive index under conditions of transparency or gain \cite{kinsler08,skaar06,Wuestner10,yavuz10,depine04,mccall02}. Utilizing the method mentioned above, the ratio between the Rabi frequencies of the two coupling beams is adjusted in order to increase chirality and produce the desired transparent or amplifying effect. Plotted in Fig.~5, both the real (a) and the imaginary (b) part of the index of refraction is shown as a function of probe field detuning for different values of this ratio. In all the cases, one obtains a negative index of refraction along with a decrease in the $n^{''}$ as the ratio between the two coupling fields is increased. For the value $\Omega_{24}/\Omega_{13}=20$ the imaginary part of the index of refraction becomes negative for a range of approximately $20\times 10^{6}\,$s$^{-1}$ over the probe field detuning signifying probe gain. This gain can be directly related to a chiral enhancement of the probe field's electric transition \cite{jung08}.

The role of LFCs in the response of the medium has also been herein investigated and previously demonstrated to play a significant role in inducing transparency \cite{friedberg89}.  The effect of local field corrections can easily be regulated by changing the atomic density, $\eta$.  
As shown in Fig.~6, as the number of atoms, and therefore the strength of the local field corrections, is increased we see a corresponding decrease in absorption as well as the expected Lorentz shift. In particular, for the value $\eta=1.5\times10^{17}\mbox{at}/\mbox{cm}^{3}$ we obtain a range of detunings for which the medium begins to amplify.


\section{Conclusions}\label{conclusions}

In this paper, a method for producing a negative index of refraction for a probe field within the infrared frequency of light is presented. A specific, practical model was demonstrated representing a dense Neon gas which has two near degenerate electric and magnetic transitions. Taken into consideration was the presence of cross polarizability terms in the material's response functions due to a strong chirality which can be utilized to manipulate the refractive index. It has been shown that by adjusting the ratio of two intense pump fields, which couple a nearly degenerate electric and magnetic transition, one can manipulate the effects of chirality in order to induce low absorption or gain in the system. In addition, local field corrections were added to the calculation where it was shown that the atomic density could also be used as a tool to further reduce absorption. The use of gases has certain potential benefits compared to metamaterials, for instance, eliminating manufacturing contraints as well as reducing experimental difficulties such as the alignment of laser and material polarizations. However, it is important to note that these two methods that have been employed for producing a negative refractive index with low absorption can be used not only in a dense atomic gas, but within similar metamaterial structures exhibiting the inherent traits of chirality and requiring LFC's\cite{Wuestner10,poutrina10}. The desired result, a consistent source of negatively refracted light, is crucial for a multitude of applications, particularly lenses with perfect resolution, subwavelength enhancement, as well as the generation transparent materials\cite{pendry00,smith04}. Although this paper has considered an isotropic, infrared source there are no restrictions to use a material of this nature and indeed further research is being undertaken to potentially show how a non-isotropic light source could be used as an additional tool in adjusting the Helmholtz equation and allowing increased control over the refractive index.  
 
%




\begin{acknowledgements}
I would like to acknowledge financial support from the Spanish Ministry of Science and Innovation under contracts FIS2008-01932-E, FIS2008-02425, and CSD2006-00019 (Consolider project ``Quantum Optical Information Technologies''), and the Catalan Government under contract SGR2009-00347. I would also like to thank Jurgen K$\ddot{\mbox{a}}$stel, Michael Fleischhauer, Jordi Mompart, Veronica Ahufinger, and Jakub Surzin for useful conversations and advice for researching and preparing this paper.
\end{acknowledgements}

\bibliographystyle{apsrev}
\bibliography{negative index}

\end{document}